\documentclass[prl,amsmath,amsfonts,amssymb,twocolumn,superscriptaddress,showpacs]{revtex4}

\usepackage{graphicx,psfrag,color}
\usepackage{dcolumn}
\usepackage{bm}
\usepackage{mathbbol}

\def\one{{\mathbb{1}}}

\begin{document}

\title{Adiabatic Perturbation Theory and Geometric Phases for Degenerate Systems}

\author{Gustavo Rigolin}
\affiliation{Department of Physics, Indiana University,
Bloomington, IN 47405, USA} \affiliation{Departamento de Fisica,
Universidade Federal de Sao Carlos, Sao Carlos, SP 13565-905,
Brazil}
\author{Gerardo Ortiz}
\affiliation{Department of Physics, Indiana University,
Bloomington, IN 47405, USA}
\date{\today}

\begin{abstract}
We introduce an adiabatic perturbation theory 
for 
quantum systems with
degenerate energy spectra. This perturbative series enables one to
rigorously establish
conditions for the validity of the adiabatic theorem of quantum mechanics
for degenerate systems.  The same formalism can be used
 to find non-adiabatic corrections to the non-Abelian Wilczek-Zee geometric
phase. 
These corrections are relevant to assess the validity of the
practical implementation of the concept of fractional exchange
statistics. We illustrate the formalism by exactly solving a
time-dependent problem and comparing its solution to the
perturbative one.
\end{abstract}

\maketitle

Controlling the dynamics of micro and nanoscopic systems is at the
heart of experimental and theoretical schemes in quantum
information and nanosciences. When, in a given physical process,
there is a clear separation of time scales into {\it fast} and
{\it slow} that fact helps in understanding its dynamics. This is the
core motivation behind the well-known adiabatic theorem of quantum
mechanics for {\it non-degenerate} systems \cite{Mes62}. But, how fast
is fast and how slow is slow? It is crucial to quantify the
validity of any time-dependent approximation in a general context
since what is at stake is the practical dynamical implementation
of concepts such as holonomic quantum computation or  the detection of
fractional (adiabatic) statistics.

A main goal of the present Letter is to asses the validity of the
adiabatic approximation/theorem for {\it degenerate} spectra by
introducing an Adiabatic  Perturbation Theory \cite{Rig08} for
Degenerate subspaces (DAPT). DAPT is a genuine perturbative
expansion in the ``velocity'' $v$ to drive the system from one
particular configuration to another, with  the perturbative
corrections to the time-dependent Schr\"odinger equation (SE)
given about the adiabatic approximation (AA). Also, DAPT is not
just a contribution to the general formalism of quantum mechanics
but also a practical and operational formulation to test the
validity of a broad range of concepts such as non-adiabatic
fractional statistics and the extent to which AA is valid for a
degenerate physical system, furnishing, as a bonus, the
corrections to AA. Finally, in the development of DAPT we also
determine corrections to the Wilczek-Zee (WZ) non-Abelian
geometric phase \cite{Wil84}.

The dynamics of a quantum system with Hamiltonian $\mathbf{H}(t)$
is dictated by the SE
\begin{equation}
\mathrm{i}\,\hbar\,v\, |\dot{\Psi}(s)\rangle =
\mathbf{H}(s) |\Psi(s)\rangle,
\label{SE2}
\end{equation}
written in terms of the rescaled time variable $s=v\,t$ ($s \in
[0,1]$), where $\hbar=h/2\pi$ with $h$ denoting Planck's constant,
$T=1/v$ is the relevant time scale of $\mathbf{H}(s)$, and the dot
means $d/ds$. Consider an explicitly time-dependent Hamiltonian
with orthonormal eigenvectors $|n^{g_n}(s)\rangle$, where
$g_n=0,1,\cdots,d_n-1$
labels states of the degenerate eigenspace $\mathcal{H}_n$ of 
dimension $d_n$ and eigenenergy $E_n(s)$, 
$\mathbf{H}(s) |n^{g_n}(s)\rangle = E_n(s) |n^{g_n}(s)\rangle,$
and assume that $d_n$ does not change during the time evolution. In
general, Eq. (\ref{SE2}) cannot be solved in closed analytic form, so
one must resort to approximation schemes. Here, we are interested in
developing a perturbative expansion, i.e., a DAPT, in the small
parameter $v$.

Inspired by the non-degenerate perturbative scheme developed in
\cite{Rig08}, we introduce the most general ansatz state $|\Psi(s)\rangle$
describing a degenerate system
\begin{equation}
|\Psi(s)\rangle = \sum_{n=0}\sum_{g_n=0}
\mathrm{e}^{-\frac{\mathrm{i}}{v}\omega_n(s)}
U_{h_ng_n}^{n}(s)b_n(s)|n^{g_n}(s)\rangle,
\label{psi}
\end{equation}
with $n=0$ representing the ground eigenspace, the dynamical phase
being given by
\begin{equation}
\omega_n(s)= \frac{1}{\hbar}\int_0^s E_n(s')\mathrm{d}s'=v \,
\omega_n(t) ,
\label{omega}
\end{equation}
and $\mathbf{U}^n(s)$ a unitary matrix whose physical meaning 
will be elucidated below. (The un-summed index $h_n$ is associated to the 
initial conditions.)
By replacing
this ansatz into (\ref{SE2}) and left  multiplying it by $\langle
m^{k_m}(s)|$ one gets
$\dot{b}_n(s)U^{n}_{h_ng_n}(s) + b_n(s)\dot{U}^{n}_{h_ng_n}(s) +
\sum_{m=0}\sum_{k_m=0}
\mathrm{e}^{-\frac{\mathrm{i}}{v}\omega_{mn}(s)} U^{m}_{h_mk_m}(s)
M^{nm}_{g_nk_m}(s)b_m(s)=0,$
with  $\omega_{mn}(s)=\omega_m(s) - \omega_n(s)$, and
$M^{nm}_{h_ng_m}(s)=\langle n^{h_n}(s)| \dot{m}^{g_m}(s)\rangle$,
which for $n\neq m$ results in
\begin{equation}
M^{nm}_{h_ng_m}(s) = \langle n^{h_n}(s)|\mathbf{\dot{H}}(s)
|m^{g_m}(s)\rangle/\Delta_{mn}(s), \label{condition}
\end{equation}
with $\Delta_{mn}(s)=E_m(s)-E_n(s)$.

Let us start analyzing the limiting situation $v \rightarrow 0$,
defining 
the {\it degenerate adiabatic approximation} (DAA). The DAA consists
in neglecting the coupling  between \textit{different} eigenspaces
$\mathcal{H}_n$ but not those within a given eigenspace, i.e., we
set
\begin{eqnarray}
M^{nm}_{g_nk_m}(s) &= \delta_{nm}M^{nn}_{g_nk_n}(s) \ \mbox{ and } \
\dot{b}_n(s)=0
\label{aprox}
\end{eqnarray}
as the {\it degenerate adiabatic condition}, implying
\begin{equation}
\dot{U}^n_{h_ng_n}(s) + \sum_{k_n=0}U^n_{h_nk_n}(s)
M^{nn}_{g_nk_n}(s)=0 \mbox{ and } \label{WZphase}
\end{equation}
\begin{equation}
\hspace*{-0.3cm} |\Psi^{(0)}(s)\rangle = \sum_{n=0}\sum_{g_n=0}
\mathrm{e}^{-\frac{\mathrm{i}}{v}\omega_n(s)}
U_{h_ng_n}^{n}(s)b_n(0)|n^{g_n}(s)\rangle . \label{psi0}
\end{equation}
In particular, if the system starts at the ground state
$|0^0(0)\rangle$ we have $b_n(0)=\delta_{0n}$ and
$U^{n}_{h_ng_n}(0)=\delta_{h_ng_n}$ leading to
\begin{equation}
|\Psi^{(0)}(s)\rangle = \sum_{g_0=0}
\mathrm{e}^{-\frac{\mathrm{i}}{v}\omega_0(s)}
U_{0g_0}^{0}(s)|0^{g_0}(s)\rangle,
\label{psi00}
\end{equation}
where we have chosen $h_0=0$ as the initial condition
$|\Psi(0)\rangle = |0^0(0)\rangle$. Similar expressions hold for
$h_n=1,2, \ldots, d_n-1$.
For $n=0$ they represent the other possible initial  conditions
$|0^1(0)\rangle, |0^2(0)\rangle, \ldots,  |0^{d_0-1}(0)\rangle$.
Had we started in a superposition of those states the condition
$U^{n}_{h_ng_n}(0)=\delta_{h_ng_n}$, or
$\mathbf{U}^{n}(0)=\mathbb{\one}$ using matrix notation,  must be
relaxed without, nevertheless, destroying the unitarity of
$\mathbf{U}^{n}(s)$. Note that for definiteness,  and without loss
of generality, we choose $h_n=h_m=0$, $\forall n,m$, in the  rest
of this Letter. Other choices would result in different initial
conditions for a fixed $\mathbf{U}^n(0)$ being, nevertheless, as
good as long  as we stick with it. As we will see below, and
hinted by Eq. (\ref{condition}),   the adiabaticity condition for
degenerate systems is connected to the gap between eigenenergies 
associated to each degenerate eigenspace.

The adiabatic phase for degenerate systems has a geometric (WZ phase)
and a dynamical component \cite{Wil84}. Here we prove how the WZ
geometric phase corresponds to the lowest-order approximation in our
DAPT.  Defining $A^{nm}_{h_ng_m}(s)=(M^{nm}_{h_ng_m}(s))^*$, Eq.
(\ref{WZphase}) can be written as $\mathbf{\dot{U}}^n(s)-
\mathbf{U}^n(s)A^{nn}(s) = 0$  (the minus sign comes from
$M^{nm}_{h_ng_m}(s)=-(M^{mn}_{g_mh_n}(s))^*$). By time integration of
this expression we arrive at the WZ phase
\begin{equation}
\mathbf{U}^{n}(s) = \mathbf{U}^{n}(0)\hat{\mathcal{T}}
\exp\left( \int_0^s\mathbf{A}^{nn}(s')ds'\right),
\end{equation}
with $\hat{\mathcal{T}}$ denoting a time-ordered exponential. 
Under a change of basis, i.e. a gauge transformation, 
$\mathbf{U}^{n}(s)$ transforms unitarily after a cyclic path 
\cite{Wil84,Kul06,Note1}. Note that
in Ref. \cite{Wil84} the authors assumed the initial state to
correspond to a single eigenvector; here we have relaxed this assumption
(cf. Eq.~(\ref{psi})), allowing us to get a stronger condition for the
validity of the DAA in its full generality, as we will see below.

We are now in a position to state the adiabatic theorem for
degenerate systems:``If the Hamiltonian of a system
$\mathbf{H}(t)$ changes slowly during the course of time, say from
$t=0$ to $t=T$, and the system is initially prepared in an
eigenstate of $\mathbf{H}$, say $|n^{g_n}(0)\rangle$,  then it
will remain in the instantaneous (snapshot) \textit{eigenspace}
$\mathcal{H}_n$ of $\mathbf{H}(t)$ during the interval $t\in
[0,T]$.'' The conditions for the validity of this theorem and the
determination of corrections to the DAA when it fails are what
concern us in the rest of the Letter. To this end, we will need to
develop a DAPT.

We want next to develop a series expansion in terms
of the small adiabatic parameter $v=1/T$, i.e., a DAPT.  As in the
non-degenerate case \cite{Rig08}, the choice of {\it ansatz} for
the state $|\Psi(s)\rangle$ is crucial for its success since one
needs to isolate the singular terms. This can be seen by noting
that the perfect ansatz would {\it factor out} the dependence of
$|\Psi(s)\rangle$ on all terms of order $\mathcal{O}(v^{0})$ and
below. In particular, terms $\mathcal{O}(v^{-1})$ and below are
the problematic ones when $v\rightarrow 0$ and should be handled
with care. Since those terms come from
$\mathrm{e}^{-\frac{\mathrm{i}}{v}\omega_n(s)}$, while the zeroth
order term comes from the WZ phase, we write down the following
ansatz
\begin{equation}
|\mathbf{\Psi}(s)\rangle =
\sum_{p=0}^{\infty}v^p|\mathbf{\Psi}^{(p)}(s)\rangle,
\;\mbox{where} \label{ansatz}
\end{equation}
\begin{equation}
|\mathbf{\Psi}^{(p)}(s)\rangle = \sum_{n=0}
\mathrm{e}^{-\frac{\mathrm{i}}{v}\omega_n(s)}
\mathbf{B}_{n}^{(p)}(s)|\mathbf{n}(s)\rangle
\;\mbox{and}\label{ansatzA}
\end{equation}
\begin{equation}
\mathbf{B}_n^{(p)}(s) = \sum_{m=0}
\mathrm{e}^{\frac{\mathrm{i}}{v}\omega_{nm}(s)}
\mathbf{B}_{mn}^{(p)}(s).
\label{ansatzB}
\end{equation}
Here $|\mathbf{\Psi}(s)\rangle$, $|\mathbf{\Psi}^{(p)}(s)\rangle$,
and $|\mathbf{n}(s)\rangle$ are $d_n$-component column vectors
while $\mathbf{B}_{n}^{(p)}(s)$ and $\mathbf{B}_{mn}^{(p)}(s)$ are
square matrices of dimensions $d_n \times d_n$, where for
definiteness we assume $d_n$ to represent the degree of degeneracy
of the most degenerate eigenspace.  For example,
%
$\mathbf{|n}(s)\rangle^{T}= \left(
\begin{array}{cccc}
|n^0(s)\rangle, & |n^1(s)\rangle, & \cdots, & |n^{d_n-1}(s)\rangle
\end{array}
\right).
$
Whenever the dimension of a given eigenspace $m$ is  smaller than
$d_n$, i.e., $d_m< d_n$, we pad zeros to those entries and to the
corresponding ones of $\mathbf{B}_{mn}^{(p)}(s)$. We will also
need the bra row vector, $\mathbf{\langle n}(s)|=
\left(\begin{array}{cccc} \langle n^0(s)|, & \langle n^1(s)|, &
\cdots, & \langle n^{d_n-1}(s)|\end{array} \right)$. Notice that
$\mathbf{\langle n}(s)|^{T}$ is a column vector which combined to
$\mathbf{|n}(s)\rangle^{T}$ by the usual matrix multiplication
rule gives $\mathbf{\langle n}(s)|^{T}\mathbf{|n}(s)\rangle^{T}$,
a $d_n \times d_n$ matrix. Each component of the vector
$|\mathbf{\Psi}(s)\rangle$ has to be interpreted as corresponding
to the evolution of an initial condition. 
Thus, if one starts at $s=0$ in the state $g_n=1$ of the ground
eigenspace $n=0$, i.e., $|{0}^1(0)\rangle$, the relevant component
describing its time evolved state is the second one, i.e.,
$[|\mathbf{\Psi}(s)\rangle]_1$. The reason for adopting this
convention is one of notational and conceptual convenience and of
mathematical simplicity in the following calculations. We will see
that this is the most general ansatz accommodating an arbitrary
initial state.

Another important ingredient \cite{Rig08} for the success of DAPT
is the setting of the right initial conditions, which amounts to
guaranteeing the following two constraints:

(i) The zeroth order must be the AA. In vector notation AA reads
(cf. Eq. (\ref{psi0})),
\begin{equation}
|\mathbf{\Psi}^{(0)}(s)\rangle = \sum_{n=0}
\mathrm{e}^{-\frac{\mathrm{i}}{v}\omega_n(s)}
b_n(0)\mathbf{U}^{n}(s)|\mathbf{n}(s)\rangle,
\label{vector0}
\end{equation}
implying $\mathbf{B}_{mn}^{(0)}(s) =
b_n(0)\mathbf{U}^{n}(s)\delta_{nm}$ and $\mathbf{B}_n^{(0)}(s) =
b_n(0)\mathbf{U}^{n}(s)$. Moreover, if we start at
$|0^0(0)\rangle$, implying $b_n(0) = \delta_{n0}$ and
$\mathbf{U}^{n}(0) = \mathbf{\one}$, we reproduce Eq. (\ref{psi00}).

(ii) For $p\geq 1$ we must have $|\mathbf{\Psi}^{(p)}(0)\rangle =
0$, implying $\mathbf{B}_{nn}^{(p)}(0) = - \sum_{m=0,m\neq
n}\mathbf{B}_{mn}^{(p)}(0)$, and thus, $\mathbf{B}_n^{(p)}(0)=0$.

Having established the initial conditions one needs to insert Eqs.
(\ref{ansatz}), (\ref{ansatzA}), and (\ref{ansatzB}), namely,
$\mathbf{|\Psi}(s)\rangle^T = \sum_{k,m=0}\sum_{p=0}^{\infty} v^p
\mathrm{e}^{-\frac{\mathrm{i}}{v}\omega_{m}(s)}
\mathbf{|k}(s)\rangle^T\mathbf{B}_{mk}^{(p)}(s)^T,$
into the SE (\ref{SE2}) in vector notation,
$\mathrm{i}\,\hbar\,v\, |\mathbf{\dot{\Psi}}(s)\rangle^T =
\mathbf{H}(s) |\mathbf{\Psi}(s)\rangle^T$, to obtain the following
recursive relation after left multiplying it with $\mathbf{\langle
n}(s)|^{T}$,
\begin{eqnarray}
\frac{\mathrm{i}}{\hbar}\Delta_{mn}(s)\mathbf{B}_{mn}^{(p+1)}(s)
=\mathbf{\dot{B}}_{mn}^{(p)}(s)
+\sum_{k=0}\mathbf{B}_{mk}^{(p)}(s)\mathbf{M}^{kn}(s) , \nonumber
\label{recursiveB}
\end{eqnarray}
where $[\mathbf{M}^{kn}(s)]_{h_kg_n}= \langle
n^{g_n}(s)|\dot{k}^{h_k}(s)\rangle = M_{g_nh_k}^{nk}(s)$ (cf. Eq.
(\ref{condition})) is the following square matrix
\begin{equation}
\mathbf{M}^{kn}(s)= \left(
\begin{array}{cccc}
M_{0,0}^{nk}& M_{1,0}^{nk}&\cdots & M_{g_{n}-1,0}^{nk} \\
M_{0,1}^{nk}& M_{1,1}^{nk}&\cdots &  M_{g_{n}-1,1}^{nk}\\
\vdots & \vdots & \ddots & \vdots \\
M_{0,g_{n}-1}^{nk} & M_{1,g_{n}-1}^{nk}& \cdots &
M_{g_{n}-1,g_{n}-1}^{nk}
\end{array}
\right).
\label{Mcal}
\end{equation}

By construction, the zeroth order term is the DAA. The case $p=0$
and $n=m$, after using the initial constraint (i), leads to
$b_n(0)(\mathbf{\dot{U}}^n(s)+\mathbf{U}^n(s)\mathbf{M}^{nn}(s))=0$,
thus recovering the expression for the WZ phase.

This recursive relation is the starting point for a DAPT.
The first non-trivial non-adiabatic contribution is first
order in $v$
\begin{eqnarray}
&&\!\!\!\!|\mathbf{\Psi}^{(1)}(s)\rangle=
\mathrm{i}\hbar \mathop{\sum_{n,m=0}}_{m \neq
n}\mathrm{e}^{-\frac{\mathrm{i}}{v}\omega_n(s)}b_{n}(0)
\mathbf{J}^{nmn}(s) \mathbf{U}^n(s)|\mathbf{n}(s)\rangle
\nonumber \\
&&-\mathrm{i}\hbar \mathop{\sum_{n,m=0}}_{m \neq n}
\mathrm{e}^{-\frac{\mathrm{i}}{v}\omega_n(s)}b_{m}(0)
\frac{_1\mathbf{W}^{mn}(0)
\mathbf{U}^n(s)}{\Delta_{nm}(0)}|\mathbf{n}(s)\rangle
\nonumber \\
&& +\mathrm{i}\hbar \mathop{\sum_{n,m=0}}_{m \neq n}
\mathrm{e}^{-\frac{\mathrm{i}}{v}\omega_m(s)}b_m(0)
\frac{\mathbf{U}^m(s)\mathbf{M}^{mn}(s)}{\Delta_{nm}(s)}
|\mathbf{n}(s)\rangle ,
\end{eqnarray}
with $_1\mathbf{W}^{mn}(s)=\mathbf{U}^m(s)\mathbf{M}^{mn}(s)
\left(\mathbf{U}^n(s)\right)^\dagger$,
\begin{equation}
\mathbf{J}^{nmn}(s) =\int_0^s\mathrm{d}s'
\left\{\frac{_2\mathbf{W}^{nmn}(s')}{\Delta_{nm}(s')}\right\},
\label{J}
\end{equation}
and
$_2\mathbf{W}^{nmn}(s)=\mathbf{U}^n(s)\mathbf{M}^{nm}(s)\mathbf{M}^{mn}(s)
\left(\mathbf{U}^n(s)\right)^\dagger$. Notice that
$|\mathbf{\Psi}^{(1)}(0)\rangle=0$ and if we have no degeneracy we
get back to the results of \cite{Rig08}.
In the particular case that the initial state is in the
ground state subspace $|\mathbf{0}(0)\rangle$
\begin{eqnarray}
&&|\mathbf{\Psi}^{(1)}(s)\rangle=
\mathrm{i}\hbar
\sum_{n=1}\mathrm{e}^{-\frac{\mathrm{i}}{v}\omega_0(s)}
\mathbf{J}^{0n0}(s) \mathbf{U}^0(s)|\mathbf{0}(s)\rangle
\nonumber \\
&&-\mathrm{i}\hbar \sum_{n=1}
\mathrm{e}^{-\frac{\mathrm{i}}{v}\omega_n(s)}
\frac{_1\mathbf{W}^{0n}(0)
\mathbf{U}^n(s)}{\Delta_{n0}(0)}|\mathbf{n}(s)\rangle
\nonumber \\
&& +\mathrm{i}\hbar \sum_{n=1}
\mathrm{e}^{-\frac{\mathrm{i}}{v}\omega_0(s)}
\frac{\mathbf{U}^0(s)\mathbf{M}^{0n}(s)}{\Delta_{n0}(s)}
|\mathbf{n}(s)\rangle .
\end{eqnarray}
More specifically, for an initial state $|0^0(0)\rangle$, the
first-order time-evolved quantum state corresponds to the  first
component $|\Psi^{(1)}(s)\rangle=[|\mathbf{\Psi}^{(1)}(s)\rangle]_0$.

Finally, we can address the validity of the general adiabatic
theorem for degenerate systems. As in the non-degenerate case, the
AA is reliable as long as  the contributions coming from the first
order correction are small. This implies the following conditions
if we start at the ground state $|0^0(0)\rangle$,
\begin{equation*}
\hbar v\left|\sum_{n=1}[\mathbf{J}^{0n0}(s)
\mathbf{U}^0(s)]_{0g_0}\right| \ll 1, \; \mbox{and for all $n\geq
1$},
\end{equation*}
\begin{eqnarray*}
\hbar v\!\left|\! \frac{[\mathbf{U}^0(s)\mathbf{M}^{0n}(s)]_{0g_n}}
{\Delta_{n0}(s)}\!-\!  \mathrm{e}^{\!-\frac{\mathrm{i}}{v}\!\omega_{n0}(\!s\!)\!}
\frac{[_1\mathbf{W}^{0n}(0)
\mathbf{U}^n(s)]_{0g_n}}{\Delta_{n0}(0)}\!\right|\!\! \ll\! 1,
\end{eqnarray*}
reducing to the ones in \cite{Rig08} when there is no degeneracy.

To illustrate our DAPT formalism, consider a four-level quantum
system subjected to a rotating classical magnetic field
$\mathbf{B}(t) = B \mathbf{r}(t)$ of constant magnitude $B$. In
spherical coordinates $\mathbf{r}(t)$ $=$ $(\sin \theta$ $\cos
\varphi(t),$ $\sin \theta$ $\sin \varphi(t),$ $\cos \theta )$,  with $0$
$\leq$ $\theta$ $\leq$ $\pi$ and $0$ $\leq$ $\varphi$ $<2\pi$ the polar and
azimuthal angles, respectively. The Hamiltonian describing the
system is \cite{Bis89}
$\mathbf{H}(t) = \frac{\hbar}{2} b\, \mathbf{r}(t) \cdot
\mathbf{\Gamma}, 
$
with $b>0$ proportional to $B$ and $\bm{\Gamma}$ $=$ $(\Gamma_x,$
$\Gamma_y,$ $\Gamma_z)$ Dirac matrices satisfying the Clifford
algebra $\{\Gamma_i,\Gamma_j\} = 2\delta_{ij}\one_4$, $j=x,y,z$.
In terms of Pauli matrices $\sigma_j$, we choose the particular
representation $\Gamma_j = \sigma_x \otimes \sigma_j$, so that
$[\Gamma_i,\Gamma_j] = 2\mathrm{i}\epsilon_{ijk}\Pi_k$, with
$\Pi_j = \one_2\otimes \sigma_j$, and $\epsilon_{ijk}$ the
Levi-Civita symbol. Then, the snapshot eigenvectors in the basis
where $\Pi_z$ is diagonal are ($n=0,1$ and
$\alpha=e^{i\varphi(t)}\sin\theta$, $\beta=\cos\theta$)
\begin{eqnarray}
\!\!\!\!\!\!\!|n^0(t)\rangle \!\!&=& \!\!
\frac{1}{\sqrt{2}}\left(\alpha^* |\!\uparrow \uparrow \rangle -\beta
|\!\uparrow\downarrow\rangle -
(-1)^n|\!\downarrow\downarrow\rangle\right),
\label{zerozerostate}\\
\!\!\!\!\!\!\!|n^1(t)\rangle \!\!&=& \!\!\frac{1}{\sqrt{2}}\left(\beta
|\! \uparrow \uparrow \rangle +\alpha |\!\uparrow\downarrow\rangle
-(-1)^n|\!\downarrow \uparrow \rangle\right), \label{zeroonestate}
\end{eqnarray}
with two-fold degenerate, time-independent, snapshot eigenvalues
$E_0=-(\hbar/2) b$ and $E_1= (\hbar/2) b$.

We can exactly solve the time-dependent SE (\ref{SE2}) when
$\varphi(t) = w\,t$,  $w>0$ representing the frequency of the
rotating magnetic field, by employing techniques similar to those
developed for the single spin-1/2 problem 
\cite{Boh93,Rab54,Note2}. Assuming that at $t=0$ the initial state
is $|0^0(0)\rangle$, the resulting time-dependent solution,
expressed in terms of the snapshot eigenvectors of
$\mathbf{H}(t)$, is
\begin{eqnarray}
\lefteqn{|\Psi(t)\rangle = e^{\mathrm{i} \frac{w t}{2}}\left [ \frac{1 +
\cos\theta}{2} A_-(t) + \frac{1 - \cos\theta}{2}A_+(t)\right ]
|0^0(t)\rangle} \nonumber \\
&+& \!\! e^{-\mathrm{i} \frac{w t}{2}}\sin\theta \ \frac{ A_+(t) -
A_-(t)}{2} |0^1(t)\rangle  \nonumber \\
&+& \!\!e^{\mathrm{i} \frac{w t}{2}}\sin^2\theta \ \frac{ B_+(t) +
B_-(t)}{2} |1^0(t)\rangle \nonumber \\
&+&\!\! e^{-\mathrm{i} \frac{w t}{2}} \sin\theta \left [ \frac{1 +
\cos\theta}{2} B_-(t) - \frac{1 - \cos\theta}{2} B_+(t) \right
]|1^1(t)\rangle, \nonumber
\end{eqnarray}
with $A_\pm(t)$ $=$ $\cos\left(\Omega_\pm t/2\right)$ $+$
$\mathrm{i}$ $\frac{b\pm w\cos\theta}{\Omega_\pm}$
$\sin\left(\Omega_\pm t/2\right)$, $B_\pm(t)$ $=$ $\mathrm{i}$
$\frac{w}{\Omega_\pm}$ $\sin(\Omega_\pm t /2)$, and $\Omega_\pm^2$
$=$ $w^2$ $+$ $b^2$ $\pm$ $2$ $w$ $b$ $\cos\theta$.

In order to develop a DAPT expansion one needs the expression for
the WZ phase. For our problem it can be determined exactly
($n=0,1$)
\begin{eqnarray}
\mathbf{U}^{n}(t) &=& \left(
\begin{array}{cccc}
U^{n}_{00}(t)& U^{n}_{01}(t) \\
U^{n}_{10}(t) & U^{n}_{11}(t)
\end{array}
\right) = \left(
\begin{array}{cccc}
z_1& -z_2^* \\
z_2 & z_1^*
\end{array}
\right) \label{WZ},
\end{eqnarray}
where $z_1=e^{\mathrm{i}wt/2}\left[
\cos\left(\frac{wt}{2}\cos\theta\right) - \mathrm{i}
\cos\theta\sin\left(\frac{wt}{2}\cos\theta\right) \right]$ and
$z_2=\mathrm{i}e^{\mathrm{i}wt/2}\sin\theta
\sin\left(\frac{wt}{2}\cos\theta\right)$. In this way, the zeroth
order term, i.e., the DAA reads
\begin{eqnarray}
|\Psi^{(0)}(t)\rangle &=&
\mathrm{e}^{-\frac{\mathrm{i}}{v}\omega_{0}(t)}
(U^{0}_{00}(t)|0^{0}(t)\rangle
+
U^{0}_{01}(t)|0^{1}(t)\rangle)\nonumber \\
&=& e^{\mathrm{i}\frac{bt}{2}} (z_1 \, |0^{0}(t)\rangle -z^*_2 \,
|0^{1}(t)\rangle ) \label{order0}
\end{eqnarray}
after replacing $\omega_0(s)=-bs/2=-bvt/2$,
which is the same obtained by expanding the exact solution
$|\Psi(t)\rangle$ up to zeroth order.
The first non-trivial correction is first order in DAPT, where the
small parameter is $v=w$. It is given by
\begin{eqnarray}
|\Psi^{(1)}(t)\rangle &=& \mathrm{i} \frac{w^2t}{4bv}\sin^2\theta
\, |\Psi^{(0)}(t)\rangle + \mathrm{i} \frac{w}{bv} \sin(bt/2)
\sin\theta \nonumber \\
&\times& \!\!\! (z_1 \sin \theta + z_2 \cos \theta) |1^0(t)\rangle
+\frac{w}{bv} [\cos(bt/2) z_2^* \nonumber \\
&+& \!\!\! \mathrm{i} \sin(bt/2) \cos\theta \, (z_1^* \sin \theta
+
z_2^* \cos \theta)]|1^1(t)\rangle,\nonumber \\
\label{order1}
\end{eqnarray}
and it \textit{exactly} coincides with the first order expansion
of the exact solution $|\Psi(t)\rangle$. Note that in working out
the expansion of the exact solution one must be aware that terms
$w^{n+1}t$ are actually order $w^n$ since $t\propto 1/v=1/w$.

We should emphasize that for most time dependent Hamiltonians one
is not able the get the exact solution. However, our DAPT is
general enough to provide a systematic expansion in powers of $v$
about the DAA for any time dependent degenerate Hamiltonian.
Furthermore, DAPT can be used to get corrections to the
non-abelian WZ phase. In our example, if the first order
correction to the DAA becomes relevant, the system's state can be
written as
$ |\Psi(s)\rangle_N = \sqrt{P(s)} |\Phi^{(0)}(s)\rangle + v
\mathrm{e}^{-\mathrm{i}bt/2}N(s)(c_0(s)|1^0(s)\rangle +
c_1(s)|1^1(s)\rangle)+ \mathcal{O}(v^2),$
where $N(s)$ normalizes the state, $P(s)=|\langle \Psi^{(0)}(s)|
\Psi(s) \rangle_N|^2$, $|\Phi^{(0)}(s)\rangle= N(s)[1 + v
f(s)]/\sqrt{P(s)} |\Psi^{(0)}(s)\rangle$, and
$f(s)=\mathrm{i}s\sin^2\theta/(4b)$. The quantities $c_0(s)$ and
$c_1(s)$ are not needed and can be obtained via
Eq.~(\ref{order1}). $P(s)$ gives the probability of measuring the
first order corrected state $|\Psi(s)\rangle_N$ in the adiabatic
state $|\Psi^{(0)}(s)\rangle$, i.e., the chances of being at
ground eigenspace. The remaining terms in $|\Phi^{(0)}(s)\rangle$
determines how $|0^0(s)\rangle$ and $|0^1(s)\rangle$ are now
superposed within the ground eigenspace. $|\Phi^{(0)}(s)\rangle$
is what is actually seen by any experiment trying to get the WZ
phase \textit{if} the first order is relevant. Keeping terms up to
first order we get $|\Phi^{(0)}(s)\rangle$ $=$ $[1 + v f(s)]
|\Psi^{(0)}(s)\rangle$ $=$ $e^{\mathrm{i}\frac{bt}{2}}$ $(x_1 \,
|0^{0}(s)\rangle + x_2 \, |0^{1}(s)\rangle )$, where $x_1=(1+v
f(s))z_1$ and $x_2=-(1+v f(s))z^*_2$ (cf. Eq. (\ref{order0})).
We can repeat the previous argument but with the system starting
at $|0^1(0)\rangle$, which gives the two remaining terms for the
corrected WZ phase (cf. the second row of Eq.~(\ref{WZ}) with
$n=0$). Putting everything together and noting that $v=w$ we get
instead of the WZ phase $\mathbf{U}^{(0)}(t)$ the following
non-abelian phase
%
$
\mathbf{V}^{(0)}(t) = \mathbf{U}^{(0)}(t) +
\frac{\mathrm{i}w^2t\sin^2\theta}{4b}\mathbf{U}^{(0)}(t).
$
%
This is what is experimentally seen if the first order correction
is relevant. Note that $\mathbf{V}^{(0)}(t)$ is unitary up to
first order and that for $v\rightarrow 0$ we recover continuously
the WZ phase. Also, the previous procedure applied to the
non-degenerate system of \cite{Rig08} gives the same correction to
the Berry phase therein computed by another method.

In conclusion, we have shown a genuine adiabatic perturbation
theory for Hamiltonians with degenerate spectrum (DAPT) as a series expansion
about the degenerate adiabatic approximation (DAA), in terms of the
``velocity" $v$ by which the system is driven away from its
initial configuration. DAPT allowed us to get a general
formulation to the adiabatic theorem for degenerate systems as
well as corrections to the non-abelian Wilczek-Zee phase when DAA
no longer holds. Finally, the key ingredients in the construction
of DAPT, differentiating it from all standard perturbation
theories and other approaches \cite{Ber87},
are three fold: (a) the rescaling of the real
time $t$ to $s=vt$, allowing a consistent perturbative expansion 
to all orders in terms of $v$; (b) the correct setting of the initial
condition at $t=0$, i.e., one should recover the zeroth order not only when
$v\rightarrow 0$ but also at $t=0$; and (c) the powerful vectorial
ansatz, Eqs.~(\ref{ansatz})-(\ref{ansatzB}), that factored out
singular terms and permitted the construction of simple matrix
recursive relations.

\begin{acknowledgments}
GR thanks the Brazilian agency CAPES for funding.
\end{acknowledgments}

\end{document}